\documentclass[useAMS,usenatbib]{mn2e}
\usepackage{amssymb}
\usepackage{natbib}
\usepackage{graphicx}
\usepackage{color}
\usepackage{lscape}
\usepackage{txfonts}
\usepackage{amsfonts}

\newcommand{\apj}{\rm ApJ}
\newcommand{\mnras}{\rm MNRAS}
\newcommand{\pasj}{\rm PASJ}

\title[]{A disc corona-jet model for the radio/X-ray
correlation in black hole X-ray binaries}
\author[Erlin Qiao and B.F. Liu]{Erlin Qiao $^{1}$\thanks{E-mail:
qiaoel@nao.cas.cn} and B.F. Liu $^{1}$\\
$^{1}$National Astronomical Observatories, Chinese Academy of
Sciences, Beijing 100012, China\\}

\begin{document}


\pagerange{\pageref{firstpage}--\pageref{lastpage}} \pubyear{}

\maketitle

\label{firstpage}

\begin{abstract}
The observed tight radio/X-ray correlation in the low spectral state
of some black hole X-ray binaries implies the strong coupling of the
accretion and jet. The correlation of $L_{\rm R} \propto L_{\rm
X}^{\sim 0.5-0.7}$ was well explained by the coupling of a
radiatively inefficient accretion flow and a jet. Recently, however,
a growing number of sources show more complicated radio/X-ray
correlations, e.g., $L_{\rm R} \propto L_{\rm X}^{\sim 1.4}$ for
$L_{\rm X}/L_{\rm Edd} \gtrsim 10^{-3}$, which is suggested to be
explained by the coupling of a radiatively efficient accretion flow
and a jet. In this work, we interpret the deviation from the initial
radio/X-ray correlation for $L_{\rm X}/L_{\rm Edd} \gtrsim 10^{-3}$
with a detailed disc corona-jet model. In this model, the disc and
corona are radiatively and dynamically coupled.  Assuming a fraction
of the matter in the accretion flow, $\eta\equiv \dot M_{\rm
jet}/\dot M$, is ejected to form the jet, we can calculate the
emergent spectrum of the disc corona-jet system.  We calculate
$L_{\rm R}$ and $L_{\rm X}$ at different $\dot M$, adjusting $\eta$
to fit the observed radio/X-ray correlation of the black hole X-ray
transient H1743-322 for $L_{\rm X}/L_{\rm Edd}> 10^{-3}$.  It is
found that always the X-ray emission is dominated by the disc corona
and the radio emission is dominated by the jet. We noted that the
value of $\eta$ for the deviated radio/X-ray correlation for $L_{\rm
X}/L_{\rm Edd} > 10^{-3}$, is systematically less than that of the
case for $L_{\rm X}/L_{\rm Edd} < 10^{-3}$, which is consistent with
the general idea that the jet is often relatively suppressed at the
high luminosity phase in black hole X-ray binaries.
\end{abstract}


\section{Introduction}
A black hole X-ray binary (BHB) is a gravitationally bound system
composed of a black hole and a normal star. BHBs are luminous in
X-rays, which is believed to be resulted by accreting the matter of
the normal star onto the black hole. According to the X-ray spectral
features and the timing properties, two typical spectral states were
identified in BHBs, i.e., the high/soft spectral state and the
low/hard spectral state (for reviews, see Tanaka \& Lewin 1995;
Remillard \& McClintock 2006). In the high/soft spectral state, the
spectrum is dominated by a peak emission around 1keV, which is
believed to be produced by a cool disc extending down to the
innermost stable circular orbit (ISCO) of a black hole (Shakura \&
Sunyaev 1973; Mitsuda et al. 1984; Belloni et al. 2000). The
spectral features of the low/hard spectral state are complicated,
which are generally thought to be produced by an inner hot accretion
flow and an outer truncated cool disc (Rees et al. 1982; Esin et al.
1997, 2001; McClintock et al. 2001; Yuan et al. 2005; Narayan \&
McClintock 2008). Recently, some observations indicate that a cool
disc may also exist in the region very close to ISCO in the low/hard
spectral state (Miller et al. 2006a, b; Tomsick et al. 2008; Reis et
al. 2010). Meanwhile, the low/hard spectral state is often
associated with the production of collimated, relativistic jets,
which are quenched in the high/soft spectral state (Fender et al.
2004).

A correlation between the radio luminosity and X-ray luminosity was
found in the low/hard spectral state of BHBs, i.e., $L_{\rm R}
\propto L_{\rm X}^{b}$, with $b \sim 0.5-0.7$ (Corbel et al. 2003,
2008, 2013; Gallo et al. 2003). The existing of the radio/X-ray
correlation for BHBs is mainly from the observations of two sources,
i.e., GX 339-4 and V404 Cyg (Corbel et al. 2013). Yuan \& Cui (2005)
interpreted this  radio/X-ray correlation, i.e., $L_{\rm R} \propto
L_{\rm X}^{\sim 0.7}$ for $L_{\rm X} \simeq 10^{-6} L_{\rm Edd}$ to
$L_{\rm X} \simeq 10^{-3} L_{\rm Edd}$ (with $L_{\rm Edd}=1.26\times
10^{38} M/M_{\odot}$ $\rm erg \ s^{-1}$ ) within the framework of a
radiatively inefficient accretion flow (RIAF)-jet model, in which a
fraction of the matter, $\eta$, in the accretion flow is assumed to
be ejected to form a jet (with $\eta \equiv \dot M_{\rm jet}/\dot
M$). In this model, the radio emission was dominated by the
self-absorbed synchrotron emission of a steady, collimated compact
jet, and the X-ray emission was dominated by a RIAF via thermal
Comptonization process. In the RIAF-jet model, if a constant $\eta$
is assumed for different $\dot M$, the radio luminosity from the jet
can be scaled as $L_{\rm R} \propto \eta \dot M^{\rm \xi}$ with $\xi
\sim 1$ (e.g., Heinz \& Sunyaev 2003), and because of the nature of
the advection in RIAF, the X-ray luminosity can be scaled as $L_{\rm
X} \propto \dot M^{q}$ with $q\sim 2$. Then, the predicted
radio/X-ray correlation is $L_{\rm R} \propto L_{\rm X}^{\xi/q}$
with $\xi/q \sim 0.5$, which is roughly consistent with
observations.  The RIAF-jet model can not only interpret the
radio/X-ray correlation, but also can explain the broadband spectral
energy distribution (SED) and most of the complex timing features of
some black hole X-ray transients, e.g., XTE J1118+480 (Yuan et al.
2005; Malzac et al. 2004). When  BHBs enter the quiescent state,
i.e., $L_{\rm X} \lesssim 10^{-6} L_{\rm Edd}$, Yuan \& Cui (2005)
predicted a steeper radio/X-ray correlation of $L_{\rm R} \propto
L_{\rm X}^{1.23}$ within the framework of the RIAF-jet model, in
which the radio emission is dominated by the jet, meanwhile the
X-ray emission is also dominated by the jet rather than  the RIAF.
Later, by fitting the SEDs of the black hole X-ray transient GRO
J1655-40 and V404 Cyg in the quiescent state, it is found that both
the radio emission and the X-ray emission can indeed be explained by
the jet (Pszota et al. 2008; Xie et al. 2014). However, due to the
very faint emission in the quiescent state, the presence of the
radio/X-ray correlation of $L_{\rm R} \propto L_{\rm X}^{1.23}$ for
$L_{\rm X} \lesssim 10^{-6} L_{\rm Edd}$ is still controversial
(Gallo et al. 2006).

We need to keep in mind the complexities of the observed radio/X-ray
correlation. For example, by critically examining the radio/X-ray
correlation in a sample of microquasars, Xue \& Cui (2007) found
that the correlation varied significantly among individual sources,
not only in terms of the shape but also of the degree of the
correlation. Recently, a statistical analysis of the data led to the
new claim of dual tracks, with some referring as the `universal'
track  and the other as being `outlier' track for $L_{\rm X}/L_{\rm
Edd} \gtrsim 10^{-3}$ (Gallo et al. 2012; Jonker et al. 2012). We
should note that the observed radio/X-ray correlation is quite
complex, and the claim of the simple two tracks is also fairly
dubious (Coriat et al. 2011). Anyhow, so far, a growing number of
sources have been discovered  with a steeper radio/X-ray
correlation, namely, $L_{\rm R} \propto L_{\rm X}^{\sim 1.4}$,
during $L_{\rm X} \gtrsim 10^{-3} L_{\rm Edd}$ [e.g., IGR
J17497-2821 (Rodriguez et al. 2007), XTE J1650-500 (Corbel et al.
2004), Swift J1753.5-0127 (Soleri et al. 2010)]. The study of the
radio/X-ray correlation of $L_{\rm R} \propto L_{\rm X}^{\sim 1.4}$
for $L_{\rm X} \gtrsim 10^{-3} L_{\rm Edd}$ is the purpose of this
work.  By collecting the archive quasi-simultaneous radio and X-ray
data between 2003 to 2010, Coriat et al. (2011) comprehensively
studied the relationship between the radio luminosity and X-ray
luminosity of the  black hole X-ray transient H1743-322. It is found
that during a high luminosity phase, corresponding to $L_{\rm
X}/L_{\rm Edd} \sim 10^{-3}$ to $10^{-1}$, H1743-322 follows the
radio/X-ray correlation with a steeper slope of $b\sim 1.4$. While
during a low luminosity phase, corresponding to $L_{\rm X}/L_{\rm
Edd} \lesssim 10^{-5}$, H1743-322 follows the radio/X-ray
correlation with a slope of $b\sim 0.6$. For $L_{\rm X}/L_{\rm Edd}$
between $10^{-5}$ to $10^{-3}$, it is probably corresponding to a
transition region between the two correlations.

The change of the radio/X-ray correlation from the low luminosity
phase to the high luminosity phase in H1743-322 may imply either the
change of the properties of the accretion flow or the change of the
different coupling in the accretion flow and jet, e.g., the change
of the dependence of the fraction of the matter $\eta$ ejected to
form the jet on the mass accretion rate $\dot M$.  The sources with
the radio/X-ray correlation of $L_{\rm R} \propto L_{\rm X}^{\sim
1.4}$ could be considered as a X-ray loud hypothesis, i.e., for a
given radio luminosity, the simultaneous X-ray luminosity of the
track with $b\sim 1.4$ is higher than that of the track with  $b
\sim 0.5-0.7$,  or in turn could be considered as a radio quiet
hypothesis, i.e., for a given X-ray luminosity, the simultaneous
radio luminosity of the track with $b\sim 1.4$ is lower than that of
the track with $b \sim 0.5-0.7$. In both hypotheses, the radio
emission is dominated by the self-absorbed synchrotron emission of
the steady, collimated compact jet, and the X-ray emission is
dominated by the accretion flow. The main difference of the two
hypotheses is the different dependence of $\eta$ on $\dot M$. In the
X-ray loud hypothesis, if a constant $\eta$ is assumed for different
$\dot M$, for a given radio luminosity, the higher X-ray luminosity
of the track with $b\sim 1.4$ is probably resulted by the transition
of the accretion flow from a RIAF to a radiatively efficiently
accretion flow, e.g, a disc corona system by Haardt \& Maraschi
(1991, 1993), Di Matteo et al. (1999), Liu et al. (2002a, 2003),
Merloni \& Fabian (2002), Cao (2009) and Huang et al. (2014), or a
luminous hot accretion flow by Yuan (2001), Xie \& Yuan (2012) and
Ma (2012). Meyer-Hofmeister \& Meyer (2014) proposed that
recondensation of gas from the corona into an inner disc can provide
additional soft photons for Comptonization, which leads to a higher
X-ray luminosity compared to the unchanged radio emission. In the
radio quiet hypothesis, a varied $\eta$  is assumed for different
$\dot M$. As suggested by Coriat et al. (2011), if there is a liner
dependence of $\eta$ on $\dot M$, i.e., $\eta \propto \dot M$, the
predicted radio/X-ray correlation should be $L_{\rm R} \propto
L_{\rm X}^{2\xi/q}$. Because for RIAF $q$ is $\sim 2$, the predicted
slope of the radio/X-ray correlation is also roughly close to the
track with $b\sim 1.4$. However, theoretically the dependence of
$\eta$ on $\dot M$ is unclear, (e.g., Pe'er \& Casella 2009),
further studies are still needed to put constraints on the relation
between $\eta$ and $\dot M$.

Observationally, there is evidence of a coupling of the hot plasma
and the jet in both BHBs and active galactic nuclei (AGNs), i.e.,
the coupling of the disc corona and jet at high mass accretion rates
and the coupling of the RIAF and jet at low mass accretion rates (Wu
et al. 2013; Zdziarski et al. 2011; Merloni et al. 2003; Falcke et
al. 2004). Meanwhile, theoretically, for $\dot M \gtrsim \alpha^2
\dot M_{\rm Edd}$ (with $\alpha$ the viscosity parameter, $\dot
M_{\rm Edd}$ = $1.39 \times 10^{18} M/M_{\rm \odot} \rm \ g\
s^{-1}$), the accretion flow will transit from a RIAF to a disc
corona system, which has been comprehensively studied by many
authors, e.g., Meyer et al. (2000a, b),  Liu et al. (2002b) and Qiao
\& Liu (2009, 2010, 2013) within the framework of the disc
evaporation model, or by Narayan \& Yi (1995b), Abramowicz et al.
(1995) and Mahadevan (1997) within the framework of the RIAF
solution.

Consequently, in this paper, we propose a disc corona-jet model to
explain the radio/X-ray correlation of $L_{\rm R} \propto L_{\rm
X}^{\sim 1.4}$ during  the high luminosity phase $L_{\rm X}/L_{\rm
Edd} \gtrsim 10^{-3}$, in which a fraction of the matter, $\eta$, in
the corona is assumed to be ejected to form the jet. So far, the
theoretical understanding of the jet formation is poor.
Specifically, it is difficult to put constraints on the dependence
of $\eta$ on $\dot M$ in our model, so we set $\eta$ as an
independent parameter on $\dot M$ to fit the observations. As an
example, by fitting the observed radio/X-ray correlation during the
high luminosity phase of black hole X-ray transient H1743-322 for
$L_{\rm X}/ L_{\rm Edd} > 10^{-3}$, we found that $\eta$ is weakly
dependent on $\dot M$, and the mean fitting result of $\eta$ is
$\sim 0.57\%$. The derived relatively smaller fitting value of
$\eta$ supports the general idea that the jet is often suppressed at
the high luminosity phase of BHBs. The disc corona-jet model is
briefly described in section \ref{model}. The numerical results are
presented in Section \ref{result}. Some comparisons with
observations are shown in Section 4. Discussions are in Section 5,
and the conclusions are in section 6.

\section{The Model}\label{model}
\subsection{Accretion flows}
The accretion flows adopted here is a geometrically thin disc
enclosed by a geometrically thick, hot corona around a central black
hole (Liu et al. 2002a, 2003). The disc is  a standard Shakura \&
Sunyaev (1973) disc, which is tightly coupled with the
plane-parallel corona. Magnetic fields are assumed to be generated
by dynamo action. As a result of Parker instability, magnetic flux
loops continuously  emerge from the disc to the corona and reconnect
with other loops. In this way, the accretion energy taken from the
thin disc is released in the corona as thermal energy and eventually
emitted away mostly in X-ray band via inverse Compton scattering.
The density of the corona is determined by an energy balance between
the downward heat conduction and  mass evaporation in the
chromospheric layer.  A detailed description of the model can be
found in Liu et al. (2002a, 2003).  The equations describing these
processes in the corona are listed as follows.
\begin{eqnarray}\label{e:energy}
{B^2 \over {4 \pi}}V_{\rm A} \thickapprox {4kT \over { m_{\rm e}
c^2}} \tau^{*} c U_{\rm rad},
\end{eqnarray}
\begin{eqnarray}\label{e:evap}
{k_{0} T^{7/2} \over \ell}  \thickapprox {\gamma \over {\gamma-1}}
nkT \bigg({kT\over m_{\rm H}}\bigg)^{1/2},
\end{eqnarray}
where  $T$ is the coronal temperature, $n$  is the coronal number
density, $B$ is the strength of magnetic field, $U_{\rm rad}$ is the
energy density of the soft photon field for Compton scattering,
$V_{\rm A}=\sqrt{B^2\over 4\pi\rho}$  is the Alfv\'en speed, $\ell$
is the length of magnetic loop, $\tau^{*}$ is the effective optical
depth, defined as $\tau^{*} \equiv  \lambda_{\rm \tau} n \sigma_{\rm
T} \ell$ with $\lambda_{\tau}\sim 1$.  Other constants have their
standard meanings. From equations (1) and (2), the temperature $T$
and density $n$ in the corona can be determined for a given magnetic
field  $B$ and radiation field $U_{\rm rad}$, and then the radiative
spectrum can be calculated out.

For a coupled disc and corona system,  the magnetic field  is
derived from an equipartition of gas pressure and magnetic pressure
in the disc, i.e., $\beta \equiv {n_{\rm disc} k T_{\rm disc}\over
{B^2/8\pi}} = 1$. The soft photons ($U_{\rm rad}$) are composed of
intrinsic disc radiation $U_{\rm rad}^{\rm in}$ and the reprocessed
radiation of backward Compton emission of the corona $U_{\rm
rad}^{\rm re}$. With energy transporting to the corona by the
magnetic field, the disc is expected to be gas pressure dominant.
Thus, the magnetic field and soft photon field can be expressed as
functions of black hole mass,  accretion rate, and distance,
\begin{eqnarray}\label{e:magg}
B=2.86\times 10^8  \alpha_{0.1}^{-{9/20}}\beta_1^{-{1/
2}}m^{-{9/20}} [\dot m_{0.1}\phi (1-f)]^{2/}r_{10}^{-{51/40}} \rm G,
\end{eqnarray}

\begin{equation}\label{e:uradin}
\begin{array}{ll}
U_{\rm rad}^{\rm in}& =aT_{\rm eff}^4=\frac{\displaystyle
4}{\displaystyle  c}\frac{\displaystyle 3G M\dot
M(1-f)\phi}{\displaystyle  8\pi
R^3} \\
& =1.14\times 10^{14} m^{-1}\dot m_{0.1}\phi (1-f)r_{10}^{-3}\ {\rm
ergs\ cm^{-3}},
\end{array}
\end{equation}

\begin{equation}\label{e:uradre}
U_{\rm rad}^{\rm re}=0.4\lambda_u U_{\rm B},
\end{equation}
where $m$, $\dot m_{0.1}$, $\alpha_{0.1}$, $\beta_1$, $r_{10}$ are
the mass of black hole, the accretion rate, the viscosity parameter,
the equipartition factor, and the distance respectively scaled by
$M_\odot$, $0.1\dot M_{\rm Edd}$, 0.1, 1 and $10R_{\rm S}$.
$\phi\equiv 1-{(R_*/R)^{1/2}}$ and $R_*=3R_{\rm S}$ is adopted as
the ISCO of a non-rotating black hole ($R_{\rm S}$= $2.95 \times
10^5 $ $M/M_{\rm \odot} \rm \ cm$). $\lambda_u$ is a factor
introduced for the evaluation of the seed field in Haardt \&
Maraschi (1991, 1993), which is around 1 in order of magnitude. Here
the energy fraction dissipated in the corona, $f$,  is not a free
parameter but can be expressed as,
\begin{equation}\label{e:f-d}
f\equiv {F_{\rm cor}\over  F_{\rm tot}}=\bigg({B^2\over
4\pi}V_A\bigg) \bigg({3GM\dot M\phi\over 8\pi R^3}\bigg)^{-1}.
\end{equation}

Combing equations (\ref{e:energy}), (\ref{e:evap}), (\ref{e:magg}),
(\ref{e:uradre}) and (\ref{e:f-d}), we get a solution of the corona
above a gas pressure-dominated disc in the case of $U_{\rm rad}^{\rm
re}>>U_{\rm rad}^{\rm in}$,

\begin{equation}\label{e:Tg2}
\begin{array}{ll}
T=&3.86 \times 10^9  \alpha_{0.1}^{-{9/80}}\beta_1^{-{1/8}}
\lambda_\tau^{-{1/4}}\lambda_u^{-{1/4}} m^{{1/80}} \\
&\times \ [\dot m_{0.1}\phi (1-f)]^{1/10} r_{10}^{-{51/160}}
\ell_{10}^{1/8}\ \rm K,
\end{array}
\end{equation}

\begin{equation}\label{e:ng2}
\begin{array}{ll}
n=&1.61\times 10^{18} \alpha_{0.1}^{-{9/40}}\beta_1^{-{1/4}}
\lambda_\tau^{-{1/2}}\lambda_u^{-{1/2}} m^{-{39/40}}\\
&\times \ [\dot m_{0.1}\phi (1-f)]^{1/5}
r_{10}^{-{51/80}}\ell_{10}^{-{3/4}}\ {\rm cm}^{-3},
\end{array}
\end{equation}

\begin{equation}\label{e:fg}
\begin{array}{ll}
f=&3.73\times 10^4(1-f)^{11/10} \alpha_{0.1}^{-{99/
80}}\beta_1^{-{11/8}} \lambda_\tau^{{1/4}}\lambda_u^{{1/4}}
m^{{11/80}}\\
&\times \ (\dot
m_{0.1}\phi)^{{1/10}}r_{10}^{-{81/160}}\ell_{10}^{3/8}.
\end{array}
\end{equation}

Given the values of the input parameters $m$, $\dot m$, $\alpha$,
$\ell$, and the initial values of $\lambda_{\tau}$ and $\lambda_u$,
we can solve equation (\ref{e:fg}) for $f$. Then $T$ and  $n$  are
solved from equations (\ref{e:Tg2}) and  (\ref{e:ng2}).  The
effective temperature of the soft photos $T_{\rm R}$

\begin{equation}
\begin{array}{ll}
\sigma T_R^4&={3GM\dot M \phi (1-f)\over 8\pi R^3}+{c\over 4}U_{\rm
rad}^{\rm re} \\
&\approx  {c\over 4} U_{\rm rad}^{\rm re}
\end{array}
\end{equation}
can also be calculated by combing equations (\ref{e:magg}) and
(\ref{e:uradre}) (with albedo always being assumed to be zero in our
calculations). With $T$, $n$ and $T_{\rm R}$, the spectra of the
disc-corona system can be calculated by Monte Carlo simulation.

In our detailed calculations, we also take into account  the
influences of  $\lambda_{\tau}$ and $\lambda_u$ and the length of
magnetic loops. Since  the corona temperature $T$, density $n$ and
energy fraction $f$ do not sensitively depend on $\lambda_{\tau}$
and $\lambda_u$ (see  equations (\ref{e:Tg2})--(\ref{e:fg})), and
the value of $\lambda_{\tau}$ and $\lambda_u$ should be $\sim1$ in
order of magnitude,  the corona spectra can not be  significantly
affected by the chosen value of $\lambda_{\tau}$ and $\lambda_u$.
Nevertheless,  we repeat the Monte Carlo simulation by adjusting the
value of $\lambda_{\tau}$ and $\lambda_u$.    We find a set of
reasonable value for $\lambda_{\tau}$ and $\lambda_u$ until the
downward-scatted luminosity is equal to the seed luminosity and the
upward-scatted luminosity is equal to the released gravitational
energy. In this way we find a self-consistent solution of the
disc-corona system and get  the corresponding emergent spectrum.  We
test the effect of the length of magnetic loop $\ell$ on the
emergent spectrum, it is found that the emergent spectrum is very
weakly dependent on $\ell$. So in our model, $\ell$ is not a free
parameter, and we set $\ell=10R_{\rm S}$ throughout the
calculations. For the detailed description of the Monte Carlo
simulation, one can also refer to Liu et al. (2003).

\subsection{Coupled disc corona-jet model}
The calculation of jet emission is based on the internal shock
scenario as described in Yuan et al. (2005). In the disc corona-jet
model, a small fraction of the matter in the accretion flow, i.e.,
$\dot M_{\rm jet}=\eta\dot M$, is assumed to be ejected to form the
jet. In this work, a conical  geometry  is considered for the jet,
and the half opening angle  of the jet is fixed at $\varphi=0.1$.
Changing the value of $\varphi$ will lead to a change of the density
of the jet, however this effect can be absorbed by the mass
accretion rate. The bulk Lorentz factor of the jet is fixed at
$\Gamma_{\rm jet}=1.2$, which is well consistent with the typically
observed value of the jet in the low/hard spectral state of BHBs
(e.g. Fender 2006). Within the jet, internal shock is produced due
to the collision of the shells with different velocities. The
internal shocks will accelerate a small fraction of the electrons to
be a power-law energy distribution with index $p$. We fix $p=2.1$ in
the calculations as suggested by Zhang et al. (2010) for fitting the
SEDs of three black hole X-ray transients J1753.5-0127, GRO J1655-40
and XTE J1720-318. The two parameters, $\epsilon_{\rm e}$ and
$\epsilon_{\rm B}$, describing the ratio of the energy of the
accelerated electrons and the amplified magnetic field to the shock
energy in the shock front are fixed at $\epsilon_{\rm e}=0.04$ and
$\epsilon_{\rm B}=0.02$ respectively, which are consistent with the
observations of gamma-ray bursts (GRB) afterglows (e.g., Panaitescu
\& Kumar 2001, 2002). Because  the Compton scattering optical depth
of the jet is small, only synchrotron emission is considered in the
calculation (Markoff et al. 2001).

\section{Numerical result}\label{result}
We calculate the emergent spectra of the disc corona-jet model
around a stellar-mass black hole with mass $M$ when the parameters
including $\dot M$, $\alpha$, and $\eta$ are specified. In this
paper, we fix the black hole mass at $M=10 M_{\odot}$, assuming a
typical viscosity parameter $\alpha=0.3$, and a constant $\eta$ for
different $\dot M$.

\begin{figure*}
\includegraphics[width=85mm,height=70mm,angle=0.0]{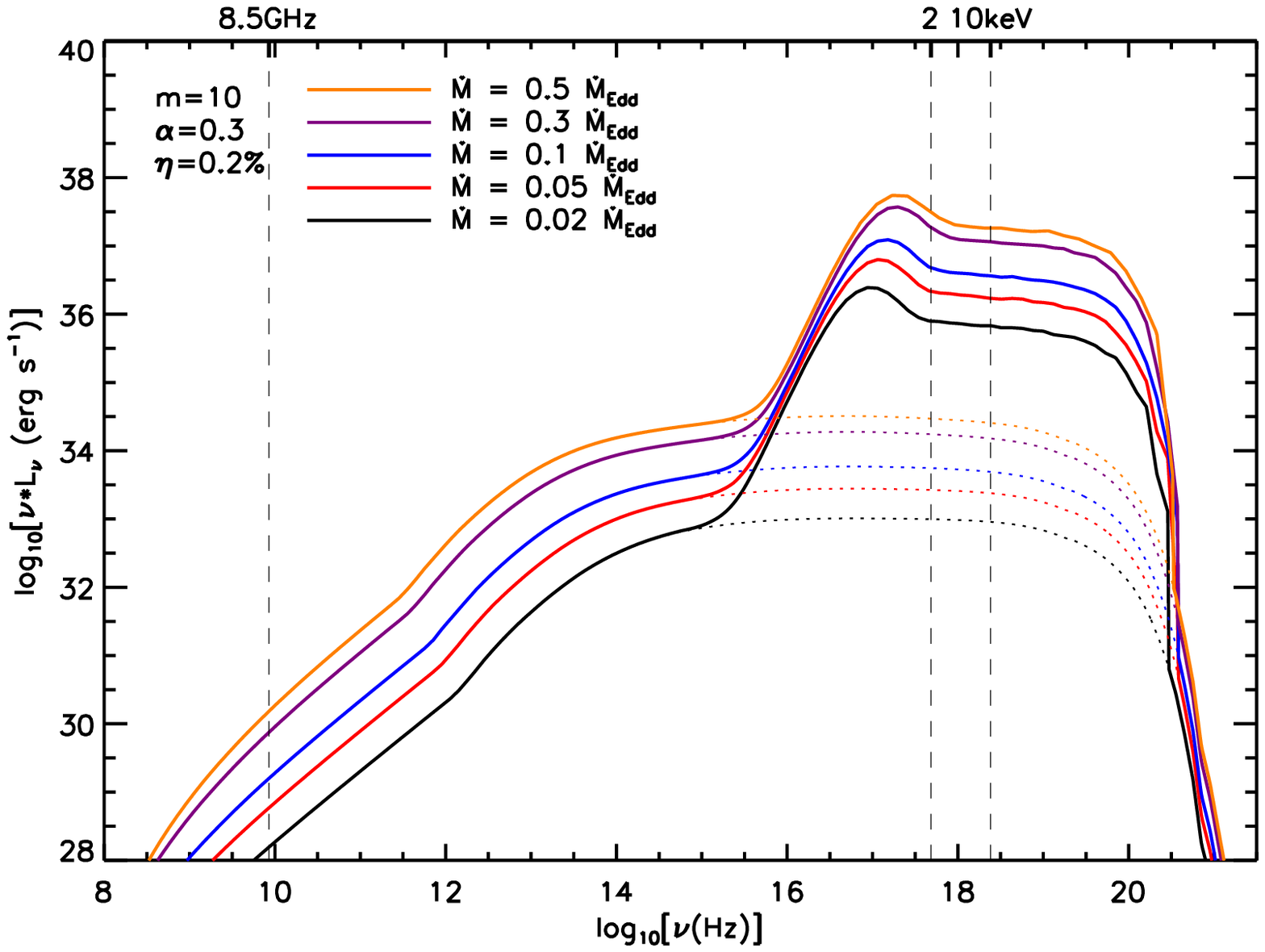}
\includegraphics[width=85mm,height=70mm,angle=0.0]{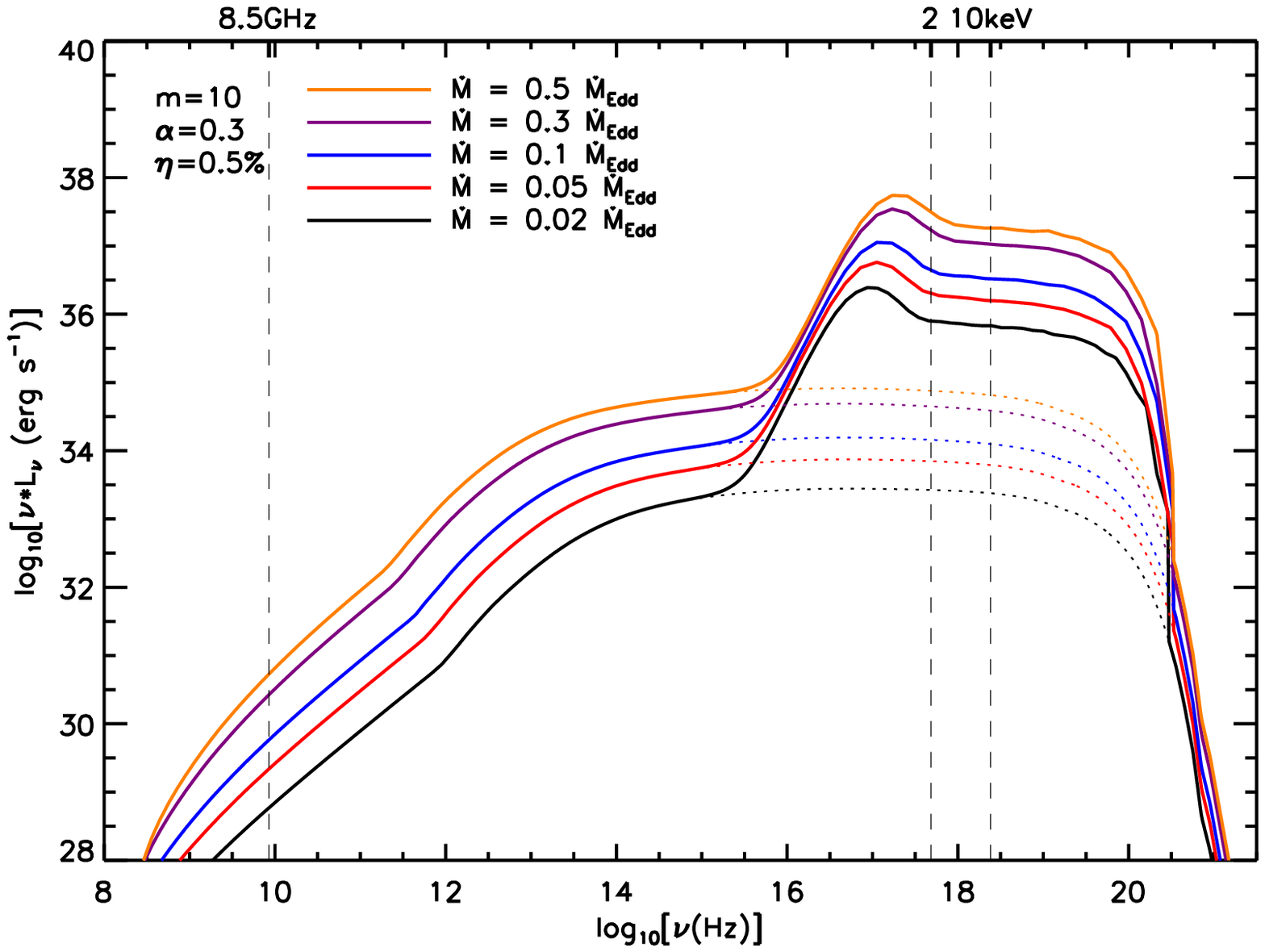}
\caption{\label{speta}} Emergent spectra of the disc corona-jet
model around a black hole with $M=10M_{\odot}$ assuming
$\alpha=0.3$. Left panel, $\eta=0.2\%$ is assumed. From the bottom
up, the solid lines are the combined emergent spectra of the disc
corona and jet for $\dot M= 0.02, 0.05, 0.1, 0.3$ and $0.5$ $\dot
M_{\rm Edd}$ respectively. The dotted lines are the emergent spectra
from the jets. Right panel: $\eta=0.5\%$ is adopted and the meaning
of the line styles are the same as in the left panel.
\end{figure*}

Given  $\eta=0.2\%$, the emergent spectra  are plotted in the left
panel of Figure \ref{speta} for different mass accretion rates,
i.e., from the bottom up $\dot M=0.02, 0.05, 0.1, 0.3$ and $0.5$
$\dot M_{\rm Edd}$. The solid lines are the total emergent spectra,
and the dotted lines are the emergent spectra from the jets. From
the left panel of Figure \ref{speta}, we can clearly see that X-ray
emission is dominated by the disc-corona system and the radio
emission is dominated by the jet 
for the mass accretion rates from $\dot M=0.02-0.5$ $\dot M_{\rm
Edd}$. Based on the emergent spectra, we calculate the radio
luminosity $L_{\rm 8.5 GHz}$ and the X-ray luminosity $L_{\rm
2-10keV}$ for different $\dot M$. The best-fitting linear regression
for the correlation between $L_{\rm 8.5 GHz}/L_{\rm Edd}$ and $\dot
M$ can be expressed as,
\begin{eqnarray}\label{fitradio}
\bigg({L_{\rm 8.5 GHz} \over L_{\rm Edd}}\bigg)=A(\eta) \bigg({\dot
M \over \dot M_{\rm Edd}}\bigg)^{\xi(\eta)},
\end{eqnarray}
where $A(\eta)\vert_{\eta=0.2\%}=10^{-8.47}$ and
$\xi(\eta)\vert_{\eta=0.2\%}$=1.42. The best-fitting linear
regression for the correlation between $L_{\rm 2-10keV}/L_{\rm Edd}$
and $\dot M$ can be expressed as,
\begin{eqnarray}\label{fitxray}
\bigg({L_{\rm 2-10keV} \over L_{\rm Edd}}\bigg)=10^{-1.23}
\bigg({\dot M \over \dot M_{\rm Edd}}\bigg)^{q},
\end{eqnarray}
where $q=1.06$, which is roughly consistent with a radiatively
efficient accretion flow with $L_{\rm X} \propto \dot M$ (e.g.
Haardt \& Maraschi 1991, 1993). Combing equations \ref{fitradio} and
\ref{fitxray}, we can derive that,
\begin{eqnarray}\label{radioxray}
\bigg({L_{\rm 8.5GHz} \over L_{\rm Edd}}\bigg)
=A(\eta)10^{1.23\xi(\eta)/q} \bigg({L_{\rm 2-10keV}\over L_{\rm
Edd}}\bigg)^{\xi(\eta)/q},
\end{eqnarray}
where $\xi(\eta)/q\vert_{\eta=0.2\%} \sim 1.35$ and
$A(\eta)10^{1.23\xi(\eta)/q}\vert_{\eta=0.2\%}=10^{-6.8}$. The
derived slope of radio/X-ray correlation from the disc corona-jet
model is steeper than that of the RIAF-jet model, i.e., for a fixed
X-ray luminosity, the increase of the radio luminosity of the
disc-corona system is more quick than that of the RIAF. One of the
reasons is that, for a given radio luminosity, due to the nature of
the high radiative efficiency, the simultaneous X-ray luminosity
predicted by the disc-corona system is intrinsically higher than
that of the RIAF. Moreover, because the X-ray luminosity of RIAF is
roughly $L_{\rm X} \propto \dot M^2$, and the X-ray luminosity of
the disc-corona system is roughly $L_{\rm X} \propto \dot M^{1.06}$,
for a fixed radio luminosity, the increase of the X-ray luminosity
of the RIAF is more quick than that of the disc-corona system, which
in turn is equivalent to the case, i.e., for a fixed  X-ray
luminosity, the increase of the radio luminosity of the disc-corona
system is more quick than that of the RIAF.

In order to check the effects of $\eta$ on the radio/X-ray
correlation, we assume another constant $\eta$  for different $\dot
M$, i.e., $\eta=0.5\%$, to calculate the emergent spectra for
comparisons with that of $\eta=0.2\%$. The emergent spectra are
shown in the right panel of Figure \ref{speta}. The best-fitting
linear regression for the correlation between $L_{\rm 8.5
GHz}/L_{\rm Edd}$ and $\dot M$ can be expressed as, ${L_{\rm 8.5
GHz} / L_{\rm Edd}}=A(\eta)({\dot M / \dot M_{\rm
Edd}})^{\xi(\eta)}$ with $A(\eta)\vert_{\eta=0.5\%}=10^{-7.94}$ and
$\xi(\eta)\vert_{\eta=0.5\%}=1.40$, and the best-fitting linear
regression for the correlation between $L_{\rm 2-10keV}/L_{\rm Edd}$
and $\dot M$ can be expressed as, ${L_{\rm 2-10keV} / L_{\rm
Edd}}=10^{-1.25}({\dot M / \dot M_{\rm Edd}})^{q}$ with $q=1.06$.
Then, the predicted radio/X-ray correlation is ${L_{\rm 8.5GHz} /
L_{\rm Edd}} =A(\eta)10^{1.25\xi(\eta)/q} ({L_{\rm 2-10keV} / L_{\rm
Edd}})^{\xi(\eta)/q}$ with
$A(\eta)10^{1.25\xi(\eta)/q}\vert_{\eta=0.5\%}=10^{-6.3}$ and
$\xi(\eta)/q\vert_{\eta=0.5\%} \sim 1.32$. Roughly, it is clearly
shown that a systematic increase of the fraction of the matter
ejected to the jet leads the radio flux to a systematic increase,
however does not change the slope of the radio/X-ray correlation
much.

In the above calculations, we assume that $\eta$  is a constant for
different $\dot M$ to produce the radio/X-ray correlation. However,
actually the dependence of $\eta$ on $\dot M$ is unclear, so $\eta$
will be set as an independent parameter on $\dot M $ to fit the
detailed  observations in the next section.

\section{Application to black hole X-ray transient H1743-322}\label{application}
H1743-322 is a X-ray transient discovered in 1977 (Kaluzienski \&
Holt 1977; Doxsey et al. 1977), then detected again by the
International Gamma-ray Astrophysics Laboratory (INTEGRAL) during
the burst in 2003. H1743-322 is well studied in both radio and X-ray
band. By comparing both the spectral features and the timing
properties of H1743-322 to the well-studied black hole X-ray
transient  XTE J1550-564, McClintock et al. (2009) argued that
H1743-322 is a accreting black hole. The black hole mass of
H1743-322 is inferred to be $\sim 10 M_{\odot}$ (e.g., Steiner et
al. 2012). Coriat et al. (2011) analyzed all the archive data of
H1743-322 observed by RXTE  between 2003 January 1 and 2010 Februay
13, meanwhile the authors conducted quasi-simultaneous ($\triangle t
\lesssim$ 1d) radio observations through Telescope Compact Array
(ATCA), Compact Array Broadband Backend (CABB), and collected the
observational data of VLA in the literature.

\begin{figure*}
\includegraphics[width=85mm,height=70mm,angle=0.0]{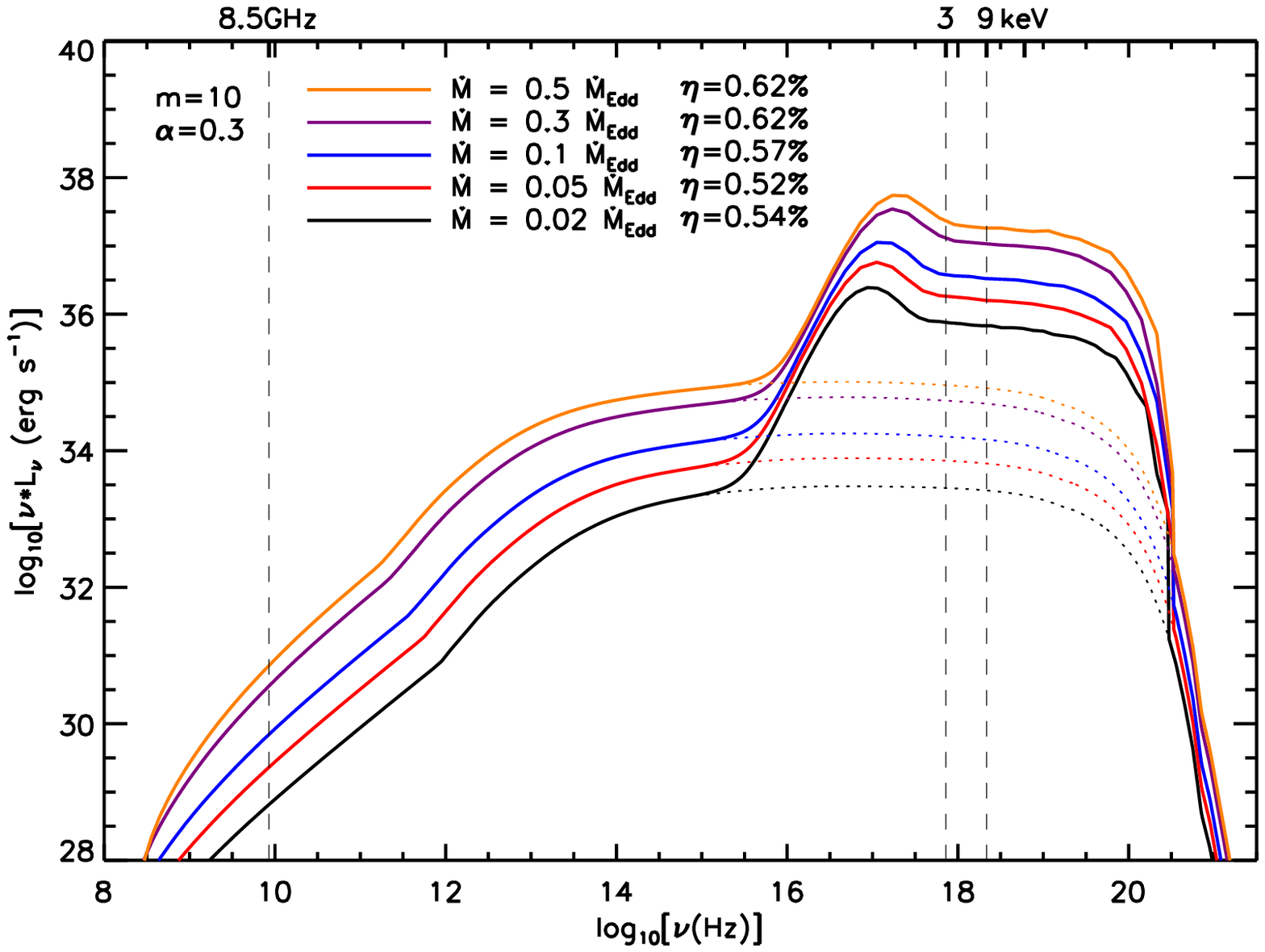}
\includegraphics[width=85mm,height=70mm,angle=0.0]{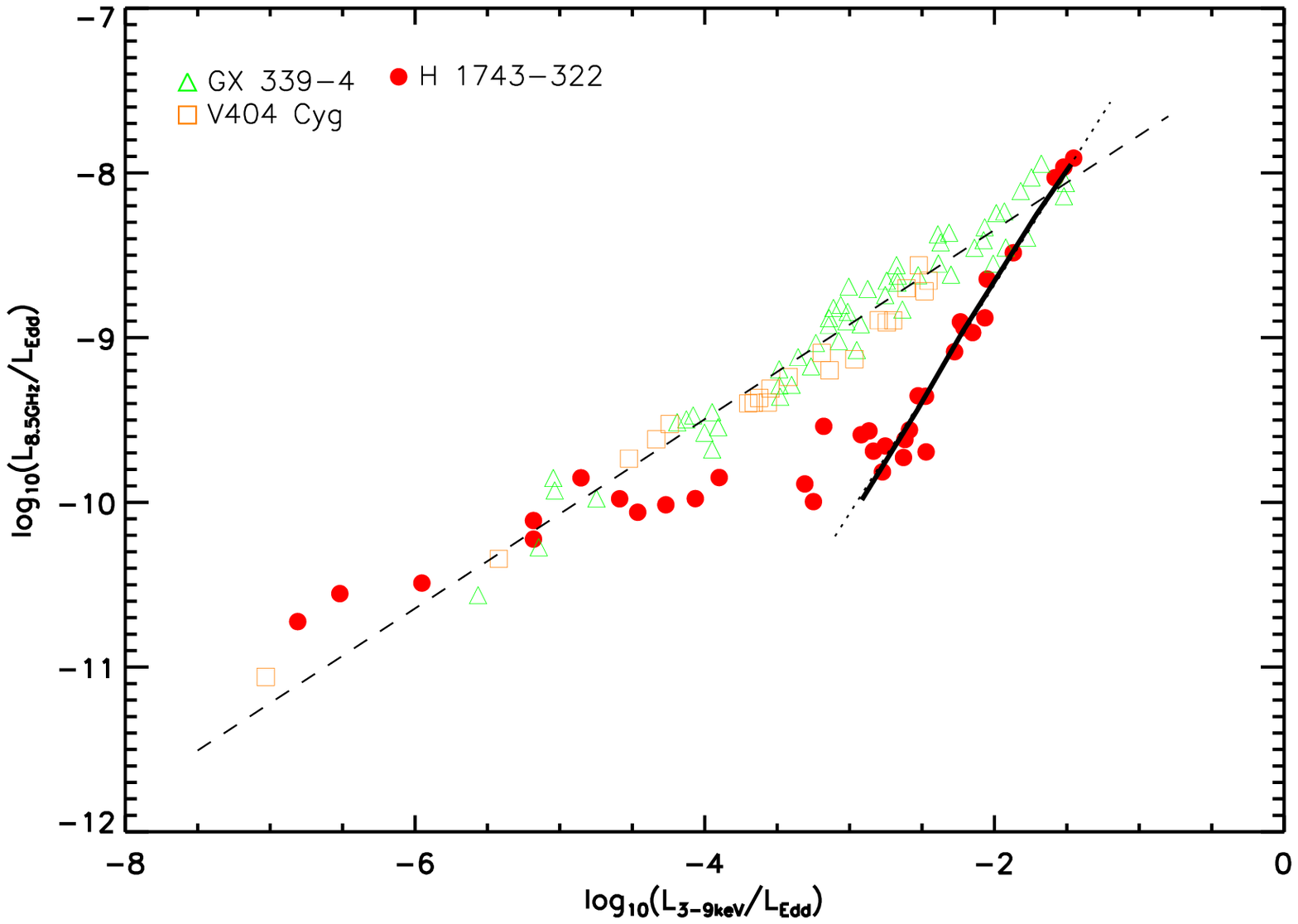}
\caption{\label{data}} Left panel: Emergent spectra of the disc
corona-jet model around a black hole with $M=10M_{\odot}$ assuming
$\alpha=0.3$ for modeling the radio/X-ray correlation of the black
hole X-ray transient H1743-322 for $L_{\rm 3-9keV}/$ $L_{\rm Edd} >
10^{-3}$. From the bottom up, the solid lines are the combined
emergent spectra of the disc corona-jet model for $\dot M= 0.02,
0.05, 0.1, 0.3$ and $0.5$ $\dot M_{\rm Edd}$, and the corresponding
dotted lines are the emergent spectra from the jet with
$\eta=0.54\%$, $0.52\%$, $0.57\%$, $0.62\%$ and $0.62\%$
respectively. Right panel: $L_{\rm 8.5 GHz}/$${L_{\rm Edd}}$ as a
function of $L_{\rm 3-9keV}/$${L_{\rm Edd}}$. The red `$\bullet$'
are the observations for H1743-322, the green $\bigtriangleup$'are
the observations for GX 339-4, and orange `$\square$' are the
observations for V404 Cyg. The dotted line is the best-fitting
linear regression of H1743-322 for $L_{\rm 3-9keV}/$ $L_{\rm Edd} >
10^{-3}$. The dashed line is the best-fitting linear regression of
GX 339-4 and V404 Cyg. The thick solid line is the model line, and
the model spectra are shown in the left panel.
\end{figure*}

Based on the data of H1743-322 from Coriat et al. (2011), we plot
the relation between $L_{\rm 3-9keV}/L_{\rm Edd}$ and $L_{\rm 8.5
GHz}/L_{\rm Edd}$ with the red sign `$\bullet$' in the right panel
of Figure \ref{data}. For $L_{\rm 3-9keV}/L_{\rm Edd} >10^{-3}$, the
best-fitting linear regression for the correlation between $L_{\rm
3-9keV}/L_{\rm Edd}$ and $L_{\rm 8.5 GHz}/L_{\rm Edd}$ is as
follows,
\begin{eqnarray}\label{datafit}
\bigg({L_{\rm 8.5 GHz} \over L_{\rm
Edd}}\bigg)=10^{-5.9}\bigg({L_{\rm 3-9keV} \over L_{\rm
Edd}}\bigg)^{1.39}.
\end{eqnarray}
This is plotted in the right panel of Figure \ref{data} with the
dotted line. Fixing $ M=10 M_{\odot}$, $\alpha=0.3$, we calculate
the radio luminosity $L_{\rm 8.5 GHz}$ and X-ray luminosity $L_{\rm
3-9keV}$ for different $\dot M$, adjusting $\eta$ to fit equation
\ref{datafit}. It is found that $\eta$ is very weakly dependent on
$\dot M$, i.e., for $\dot M=0.02$ $\dot M_{\rm Edd}$, $\eta=0.54\%$;
for $\dot M=0.05$ $\dot M_{\rm Edd}$, $\eta=0.52\%$; for $\dot
M=0.1$ $\dot M_{\rm Edd}$, $\eta=0.57\%$; for $\dot M=0.3$ $\dot
M_{\rm Edd}$, $\eta=0.62\%$; for $\dot M=0.5$ $\dot M_{\rm Edd}$,
$\eta=0.62\%$. The  mean fitting value of $\eta$ is $\sim 0.57\%$.
The thick solid line in the right panel of Figure \ref{data} is the
model line. The corresponding emergent spectra with mass accretion
rates are plotted in the left panel of Figure \ref{data}.  Yuan \&
Cui (2005) fitted the radio/X-ray correlation of $L_{\rm R} \propto
L_{\rm X}^{\sim 0.7}$ for $L_{\rm X} \simeq 10^{-6} L_{\rm Edd}$ to
$L_{\rm X} \simeq 10^{-3} L_{\rm Edd}$ within the framework of the
RIAF-jet model. They found that the fitting result was very
sensitive to a parameter $\delta$, which denotes the fraction of the
viscosity dissipated energy directly heating the electrons in the
RIAF. For $\delta=0.5$, they found that $\eta$ was highly dependent
on $\dot M$, decreasing from $\eta \sim 10\%$ to $\eta \sim 1\%$
with mass accretion rate from $\dot M \sim 10^{-3} \dot M_{\rm Edd}$
to $\sim 10^{-2} \dot M_{\rm Edd}$. For a smaller value of
$\delta=0.01$, they found that $\eta$ was nearly a constant with
$\dot M$, i.e., $\eta \sim 1\%$. Indeed, we have tested the value of
$\eta$ for H1743-322 during the phase $L_{\rm X}/L_{\rm Edd}
\lesssim 10^{-3}$ within the framework of RIAF-jet model. We found
$\eta \sim 10\%$. It is clear that the fitting value of $\eta$ from
the RIAF-jet model for the low luminosity phase is systematically
higher than that of the fitting value of $\eta$ from the disc
corona-jet model for the high luminosity phase, which is consistent
with the general idea that the jet is often relatively suppressed at
the high luminosity phase in BHBs (Fender 2004).

\begin{figure*}
\includegraphics[width=85mm,height=70mm,angle=0.0]{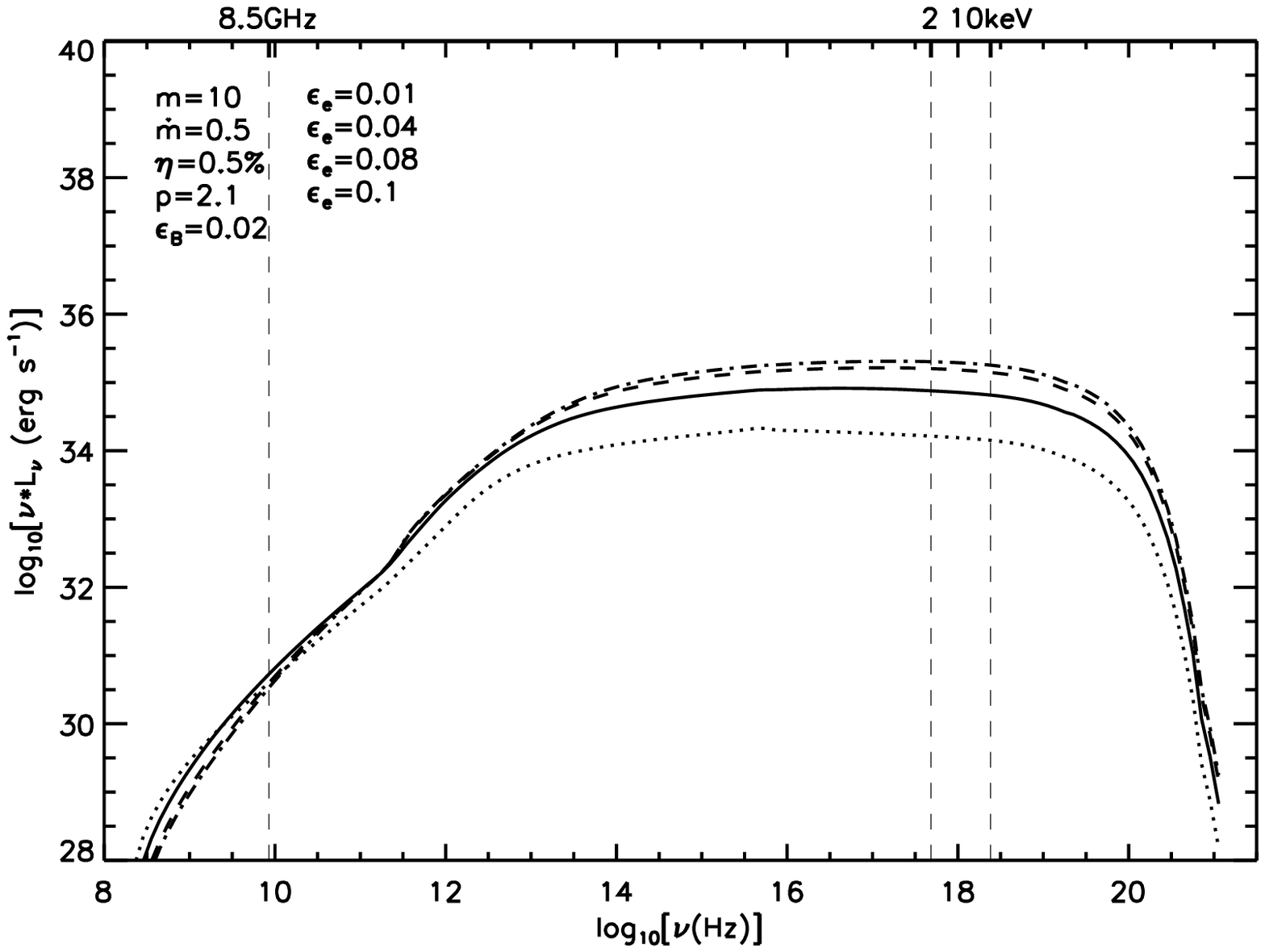}
\includegraphics[width=85mm,height=70mm,angle=0.0]{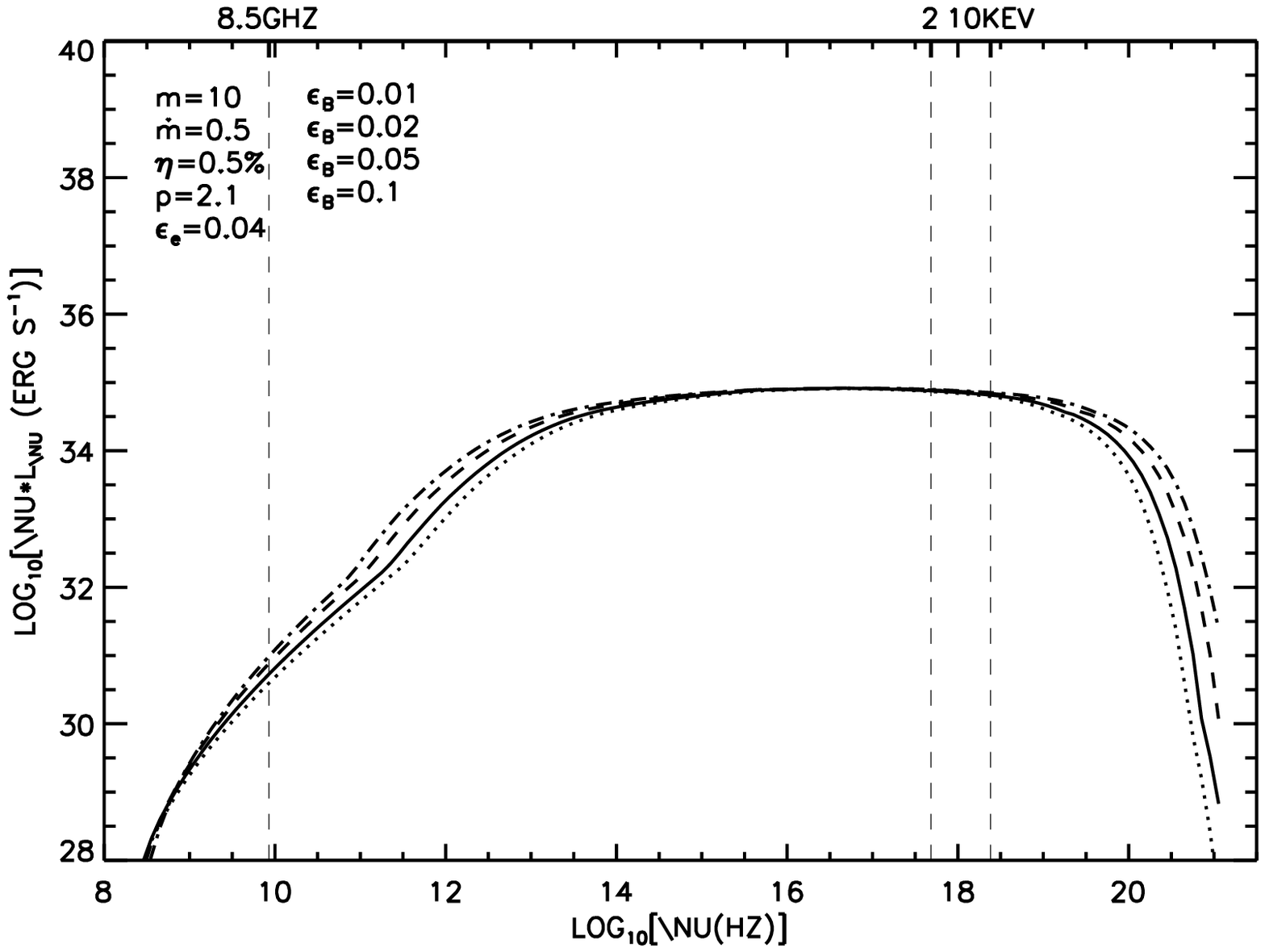}
\caption{\label{jet}}Emergent spectra of the jet for different
parameters. In the left panel, $M=10M_{\odot}$, $\dot M=0.5 \dot
M_{\rm Edd}$, $\eta=0.5\%$, $p=2.1$ and $\epsilon_{\rm B}=0.02$ are
fixed, and from the bottom up, $\epsilon_{\rm e}$ are $0.01$,
$0.04$, $0.08$ and $0.1$ respectively. In the right panel,
$M=10M_{\odot}$, $\dot M=0.5 \dot M_{\rm Edd}$, $\eta=0.5\%$,
$p=2.1$ and $\epsilon_{\rm e}=0.04$ are fixed, and from the bottom
up, $\epsilon_{\rm B}$ are $0.01$, $0.02$, $0.05$ and $0.1$
respectively.
\end{figure*}

In the present jet model, besides the mass rate in the jet $\dot
M_{\rm jet}$, there are still three parameters, which can affect the
emission of the jet, i.e., the power-law energy distribution index
of the accelerated electrons $p$, the ratio of the energy of the
accelerated electrons and the amplified magnetic field to the shock
energy, $\epsilon_{\rm e}$ and $\epsilon_{\rm B}$. In this work,
$p=2.1$, $\epsilon_{\rm e}=0.04$ and $\epsilon_{\rm B}=0.02$ are
fixed respectively. By modeling the broadband emission of the
afterglow of eight GRBs, Panaitescu \& Kumar (2001) derived that
$p=1.87\pm0.51$, $\epsilon_{\rm e}=0.062\pm 0.045$ and $\rm log
\epsilon_{\rm B}=-2.4\pm 1.2$. Yuan et al. (2005) fitted the
multiwavelength observations of XTE J1118+480 with RIAF-jet model,
in which the best fitting results are $p=2.24$, $\epsilon_{\rm
e}=0.06$ and $\epsilon_{\rm B}=0.02$ respectively. Based on RIAF-jet
model, Zhang et al. (2010) fitted the simultaneous multiwavelength
observations of three black hole X-ray transients J1753.5-0127, GRO
J1655-40, and XTE J1720-318. It is found that the best fitting
results in J1753.5-0127 are $p=2.1$, $\epsilon_{\rm e}=0.04$ and
$\epsilon_{\rm B}=0.02$, in GRO J1655-40 are $p=2.1$, $\epsilon_{\rm
e}=0.06$ and $\epsilon_{\rm B}=0.02$, and in XTE J1720-318 are
$p=2.1$, $\epsilon_{\rm e}=0.06$ and $\epsilon_{\rm B}=0.08$. It has
been tested that the different values of $p$ has very minor effects
on the radio emission, however has significantly effects on the
X-ray emission ($\rm \ddot{O}$zel et al. 2000; Figure 4 in Yuan et
al. 2003). Due to the X-ray emission is always dominated by the
emission of disc and corona, the change of $p$ in an observationally
reasonable range will not change our results. Throughout the
calculation $p=2.1$ is fixed.  We test the effects of $\epsilon_{\rm
e}$ and $\epsilon_{\rm B}$ on the jet emission. In the left panel of
Figure \ref{jet}, fixing $M=10 M_{\odot}$, $\alpha=0.3$, $\dot
M=0.5$, $\eta=0.5\%$, $p=2.1$ and $\epsilon_{\rm B}=0.02$, we plot
the emergent spectra of jet for $\epsilon_{\rm e}=0.01, 0.04, 0.08,
0.1$ respectively. It is found that, $\epsilon_{\rm e}$  is
sensitive to the X-ray emission, and insensitive to the radio
emission. However, in the observationally inferred range of
$\epsilon_{\rm e} \sim 0.01-0.1$, it is clear that the X-ray
emission is still dominated by the disc and corona instead of the
jet, so the change of $\epsilon_{\rm e}$ will not change our results
(see the right panel of Figure \ref{speta} for comparison).
Throughout the calculation $\epsilon_{\rm e}=0.04$ is fixed. In the
right panel of Figure \ref{jet}, fixing $M=10 M_{\odot}$,
$\alpha=0.3$, $\dot M=0.5$, $\eta=0.5\%$, $p=2.1$ and $\epsilon_{\rm
e}=0.04$, we plot the emergent spectra of jet for $\epsilon_{\rm
B}=0.01, 0.02, 0.05, 0.1$ respectively. It is found that, increasing
the value of $\epsilon_{\rm B}$ will increase the radio emission,
and nearly does not change the X-ray emission. As discussed in
Section 3.1, an increase of $\dot M_{\rm jet}$ will also increase
the radio emission. So detailed spectral fitting are needed to
disentangle the effects of $\epsilon_{\rm B}$ and $\dot M_{\rm jet}$
on the jet emission, which is beyond the study of the present paper.
In order to compare with other works for the fraction of the matter
ejected to form the jet in the low luminosity phase (e.g, Yuan \&
Cui 2005), we take the same value of $\epsilon_{\rm B}=0.02$ as
adopted in Yuan \& Cui (2005). In a word, although the parameters in
the jet are uncertain, we still can conclude that the jet is
relatively suppressed at the high luminosity phase in BHBs.

The mechanism of jet formation is unclear. Narayan \& McClintock
(2012) collected a small sample composed of five black hole X-ray
transients with precise spin measurements. Meanwhile the authors
estimated the ballistic jet power using the data at 5$\rm GHz$ radio
observations. It is found that the estimated jet power is correlated
with the square of $a_{*}$ (with $a_{*}=cJ/GM^{2}$, $J$ is angular
momentum of the black hole), which is very close to the theoretical
scaling derived by Blandford \& Znajek (1977). However, by
separately considering  the ballistic jet and the hard steady jet,
Fender, Gallo \& Russell (2010) found that there is no evidence for
the correlation between the jet power and the black hole spin.
Recently, an interesting paper argued that the jet formation may be
correlated with the hot plasma, namely, the jet power is correlated
with the RIAF when the Eddington ratio is less than $\sim 1\%$ and
the jet power is correlated with the hot corona above the cool disc
when the Eddington ratio is greater than $\sim 1\%$ (Wu et al.
2013). Our study may put some constrains on the mechanism of jet
formation, i.e.,  by suggesting that the relative strength of the
jet power may be inversely correlated with the Eddington ratio in an
accreting black hole (e.g., K$\ddot{\rm o}$rding, Falcke, \& Markoff
2002; Fender, Gallo, \& Jonker 2003).

As a comparison, in the right panel of Figure \ref{data}, we also
plot the relation between $L_{\rm 3-9keV}/L_{\rm Edd}$ and $L_{\rm
8.5 GHz}/L_{\rm Edd}$ for GX 339-4 with the sign of green
`$\bigtriangleup$', and V404 Cyg with the sign of orange
`$\square$'.  For $L_{\rm X}>10^{-3} L_{\rm Edd}$, the radio/X-ray
correlation with $L_{\rm 8.5GHz} \propto L_{\rm 3-9keV}^{\sim 0.6}$
is expected to be interpreted within the framework of the RIAF-jet
model, as suggested for the black hole X-ray transient XTE
J1118+480. However, how to justify the existence of the RIAF at the
high luminosity phase? Qiao \& Liu (2009) studied the effect of the
viscosity parameter $\alpha$ on the critical mass accretion rate for
the transition from a RIAF to a disc corona system within the
framework of the disc evaporation model. The authors derived that
$\dot M_{\rm crit} \propto \alpha^{2.34}$ $\dot M_{\rm Edd}$, i.e.,
a larger value of $\alpha$ can increase the critical mass accretion
rate for the existence of the RIAF.  A similar result to the
critical mass accretion rate for the existence of the RIAF, i.e.,
$\dot M_{\rm crit} \propto \alpha^{2}$ $\dot M_{\rm Edd}$ was also
derived from the RIAF solution (Narayan \& Yi 1995b; Mahadevan
1997). By summarizing the observational data of dwarf nova
outbursts, outbursts of X-ray transients, and variability in AGN and
so on, King et al. (2007) inferred that the value of $\alpha$ is in
the range of $\sim 0.1-0.4$. Extremely, Narayan (1996) took
$\alpha=1$ to explain the BHB systems with high transition
luminosities. Theoretical understandings for the viscosity have been
studied for many years since the pioneering work by Shakura \&
Sunyaev (1973) (Balbus \& Hawley 1991; Hawley et al. 1995; Hawley \&
Krolik 2001). The estimated value of $\alpha$ from numerical
simulations is roughly one order of magnitude smaller than that of
the observations (Stone 1996; Hirose et al. 2006). However, indeed
there are numerical simulations showing that the value of $\alpha$
sometimes even can exceed unity in the corona (e.g., Machida et al.
2000).

\section{Discussion}
In this work, we proposed a disc corona-jet model to explain the
observed radio/X-ray correlation of $L_{\rm R} \propto L_{\rm
X}^{\sim 1.4}$ for $L_{\rm X}/L_{\rm Edd} \gtrsim 10^{-3}$ in BHBs.
We noted that a similar disc corona-jet model was also proposed for
explaining this radio/X-ray correlation, in which the X-ray emission
is also dominated by the disc and corona, and the radio emission is
dominated the jet (Huang et al. 2014). Both in our work and Huang et
al. (2014), for the disc-corona model, it is assumed that the
magnetic field is generated by dynamo action in the accretion disc,
then due to buoyancy, the magnetic loops emerge from the accretion
disc into the corona and reconnect with other loops. In this way the
accretion energy is released in the corona as thermal energy and
eventually emitted away via inverse Compton scattering. By studying
the energy balance of the disc and the corona, Huang et al. (2014)
solved the structure of the cold accretion disc. However, they can
not self-consistently determine the temperature of the corona $T$
and the Compton scattering optical depth $\tau$, which are two very
important quantities for determining the shape of the Compton
emergent spectrum. Theoretically, $T$ and $\tau$  should be solved
independently for determining the Compton emergent spectrum. In
Huang et al (2014), it is argued that the observationally inferred
value of $\tau$ is in the range of $\sim 0.1-0.8$, then through
their calculation, $\tau=0.5$ is fixed to solve $T$ for fitting this
radio/X-ray correlation. In our model, we do not fix the value of
$\tau$. We performed self-consistent Monte carlo simulation to treat
the structure of the disc-corona system. The fraction of the
dissipated energy in the corona $f$, the temperature of the corona
$T$ and the Compton scattering optical depth $\tau$ can be
self-consistently determined, which are important advantages of our
model.

We note that in our model, currently in order to simplify the
calculation of the complex interaction between the disc and corona,
we always set the albedo, i.e., `a=0', to conduct the Monte Carlo
simulation, which means that the irradiation photons from the corona
are fully absorbed by the accretion disc, then are reprocessed as
the soft photons for the inverse Compton scattering in the corona.
Since albedo `a=0' is adopted, we don't have reflection component in
the emergent spectra. We should keep in mind that in our model, in
the gas pressure dominated case, the origin of soft photons for the
inverse Compton scattering is dominated by the reprocessed soft
photons rather than the intrinsic soft photons of the accretion disc
itself. A change the value of the albedo `a', e.g., an increase the
value of `a' means the soft photon luminosity caused by reprocess
decreases, so the Compton luminosity decreases, meanwhile the
reflection luminosity increases. In our Monte Carlo simulation, for
the energy conservation, we always set that the emergent luminosity
is equal to the released gravitational energy. So a change the value
of the albedo `a' in a reasonable range will change the shape of the
X-ray spectrum (relative strength of the Compton component and the
reflection component), but will slightly change the X-ray
luminosity. So the effects of the albedo on the emergent spectrum
have only very little change to our results.

As we know, the observed high frequency quasi-periodic oscillation
(HFQPO), e.g., 150-450 Hz, are consistent with the Keplerian
frequency near the ISCO of a Schwarzschild black hole with masses
15-5 solar masses.  Meanwhile, the observed HFQPOs do not change
significantly despite the sizable change in the X-ray luminosity,
suggesting the connections between the HFQPOs and the mass and spin
of the black hole.  If the black hole mass is well constrained, the
HFQPOs can be used to measure the spin of the black hole. The
observed pairs of frequencies in a 3:2 ratio suggest that the HFQPOs
are probably produced by some types of resonance mechanism (e.g.,
Abramowicz \& Kluzniak 2001). In the present disc-corona model, for
simplicity, we assume the accretion disc always extending down to
the ISCO of a Schwarzschild black hole, i.e, ISCO is fixed at
3$R_{\rm S}$. The incorporation of the effects of the spin to the
disc-corona model is necessary in the future to make the model more
realistic, meanwhile to match the observed HFQPOs.

Observationally, there is a positive correlation between the
Eddington ratio $\lambda$ and hard X-ray index $\Gamma$ for $\lambda
\gtrsim 0.01$ (Wu \& Gu 2008, Qiao \& Liu 2013). In the present
disc-corona model, the hard X-ray photon index is $\Gamma \sim 2.1$,
and does not change with the mass accretion rates. In the present
paper, for simplicity, we only consider the gas-pressure dominated
case for the structure of the accretion disc, i.e., only the
hard-state solution is considered (Liu et al. 2002; 2003). As
discussed in Liu et al. (2003), the gas-pressure dominated accretion
disc can exist for all the mass accretion rates. When the system is
accreting at $\dot M > 1.2 \dot M_{\rm Edd}$, the radiation-pressure
dominated accretion disc can exist extending to $50R_{\rm S}$, which
predicts a soft-state solution. When the accretion rate is at $0.3
\dot M_{\rm Edd} \lesssim \dot M \lesssim 1.2M_{\rm Edd}$, the
accreting system can be at a state between the hard state and the
soft state with hard X-ray index varying with mass accretion rates.

Meanwhile, in Huang et al. (2014), the jet power is  estimated
according to the proposed hybrid jet model (Meier 1999, 2000), from
which the radio luminosity is estimated based on the empirical
relation from Cyg X-1 and GRS 1915+105 (Falcke \& Biermann 1996;
Heinz \& Sunyeav 2003; Heinz \& Grimm 2005). In the present paper,
based on the internal shock scenario, we calculated the emergent
spectrum of the jet, which makes our results more easy to compare
with observations.

\section{Conclusion}
In this work, we investigate the radio/X-ray correlation of $L_{\rm
R} \propto L_{\rm X}^{\sim 1.4}$ for $L_{\rm X}/L_{\rm Edd} \gtrsim
10^{-3}$  within the framework of a disc corona-jet model, in which
a fraction of the matter, $\eta$, is assumed to be ejected to form
the jet.  We calculate the slope of the radio/X-ray correlation by
assuming a constant $\eta$ for different $\dot M$. For $\eta=0.2\%$,
it is found that $L_{\rm 8.5GHz} \propto L_{\rm 2-10keV}^{\xi/q}$
with $\xi/q \sim 1.35$. For $\eta=0.5\%$, we derive that $L_{\rm
8.5GHz} \propto L_{\rm 2-10keV}^{\xi/q}$ with $\xi/q \sim 1.32$,
which is very close to the case of $\eta=0.2\%$. As an example, for
different $\dot M$, by changing the value of $\eta$, we fit the
observed radio/X-ray correlation of black hole X-ray transients
H1743-322 for $L_{\rm 3-9keV}/L_{\rm Edd} > 10^{-3}$. It is found
that $\eta$ is weakly dependent on $\dot M$, and the mean fitting
value of $\eta$ is $\sim 0.57\%$. We note an interesting result,
i.e., the mean fitting result of $\eta\sim 0.57\%$ for the
radio/X-ray correlation of $L_{\rm R} \propto L_{\rm X}^{\sim 1.4}$
during the high luminosity phase is systematically less than that of
the case for the low luminosity phase (at least $\eta \gtrsim 1\%$),
which may put some constraints on the jet formation, i.e., by
suggesting that the strength of the jet power is relatively
suppressed during the high luminosity phase in BHBs.

\section*{Acknowledgments}
E.L. Qiao thanks the jet code from Dr. Hui Zhang. We thank the
referee for his/her very useful suggestions and comments, especially
the expert view on the observational aspects to the manuscript. E.L.
Qiao appreciates Dr. Francesca Panessa for the English correction.
We thank the very useful discussions with Prof. Weimin Yuan. This
work is supported by the National Natural Science Foundation of
China (grants 11303046, 11033007, 11173029, and U1231203) and the
Strategic Priority Research Program ¡±The Emergence of Cosmo-logical
Structures¡± of the Chinese Academy of Sciences (Grant No.
XDB09000000).

\label{lastpage}

\end{document}